\documentclass{article}

\usepackage{arxiv}
\usepackage[acronym]{glossaries}
\usepackage[utf8]{inputenc} 
\usepackage[T1]{fontenc}    
\usepackage{hyperref}       
\usepackage{url}            
\usepackage{booktabs}       
\usepackage{amsfonts}       
\usepackage{nicefrac}       
\usepackage{microtype}      
\usepackage{lipsum}
\usepackage{graphicx}
\graphicspath{ {./images/} }

\makeglossaries
\newacronym{cnn}{CNN}{Convolutional Neural network}
\newacronym{frcnn}{FRCNN}{Faster Region-based Convolutional Neural Networks}
\newacronym{rpn}{RPN}{Region Proposal Network}
\newacronym{svm}{SVM}{Support Vector Machine}
\newacronym{lsf}{LSF}{Low Spatial Frequency}
\newacronym{hsf}{HSF}{High Spatial Frequencies}
\newacronym{meg}{MEG}{Magneto Encephalography}
\newacronym{fmri}{fMRI}{functional Magnetic Resonance Imaging}
\newacronym{eeg}{EEG}{Electroencephalography}
\newacronym{erp}{ERP}{Event-Related brain Potential}
\newacronym{frn}{FRN}{Feedback Related Negativity}
\newacronym{ofc}{OFC}{Orbitofrontal Cortex}
\newacronym{pfc}{PFC}{Prefrontal Cortex}
\newacronym{itc}{ITC}{Inferior Temporal Cortex}

\title{Biologically Inspired Visual System Architecture for Object Recognition in Autonomous Systems}

\author{
 Dan Malowany \\
  Department of Electrical and Computer Engineering\\
  Ben-Gurion University of the Negev\\
  Israel \\
  \texttt{danyosef@post.bgu.ac.il} \\
   \And
 Hugo Guterman \\
  Department of Electrical and Computer Engineering\\
  Ben-Gurion University of the Negev\\
  Israel \\
  \texttt{hugo@bgu.ac.il} \\
}

\begin{document}
\maketitle
\begin{abstract}
Computer vision is currently one of the most exciting and rapidly evolving fields of science, which affects numerous industries. Research and development breakthroughs, mainly in the field of convolutional neural networks, opened the way to unprecedented sensitivity and precision in object detection and recognition tasks. Nevertheless, the findings in recent years on the sensitivity of neural networks to additive noise, light conditions and to the wholeness of the training dataset, indicate that this technology still lacks the robustness needed for the autonomous robotic industry.
In an attempt to bring computer vision algorithms closer to the capabilities of a human operator, the mechanisms of the human visual system was analyzed in this work. Recent studies show that the mechanisms behind the recognition process in the human brain include continuous generation of predictions based on prior knowledge of the world. These predictions enable rapid generation of contextual hypotheses that bias the outcome of the recognition process. This mechanism is especially advantageous in situations of uncertainty, when visual input is ambiguous. In addition, the human visual system continuously updates its knowledge about the world based on the gaps between its prediction and the visual feedback.
Convolutional neural networks are feed forward in nature and lack such top-down contextual attenuation mechanisms. As a result, although they process massive amounts of visual information during their operation, the information is not transformed into knowledge that can be used to generate contextual predictions and improve their performance. In this work, an architecture was designed that aims to integrate the concepts behind the top-down prediction and learning processes of the human visual system with the state of the art bottom-up object recognition models, e.g., deep convolutional neural networks. The work focuses on two mechanisms of the human visual system: anticipation-driven perception and reinforcement-driven learning. Imitating these top-down mechanisms, together with the state of the art bottom-up feed-forward algorithms, resulted in an accurate, robust, and continuously improving target recognition model.
\end{abstract}


\section{Introduction}
\label{intro}
While there are many technological gaps that inhibit the development of the autonomous systems field, there is no doubt that a significant factor in this delay is that it has been much harder than expected to give robotic agents the capabilities to analyze their ever-changing environment, detect and classify the objects surrounding them, and interpret the interaction between them. Visual object recognition is an extremely difficult computational task. The core problem is that each object in the world can cast an infinite number of different 2D images onto the retina as the object's position (translation and rotation), pose, lighting, and background vary relative to the viewer.

The state of the art computer vision algorithms, \acrfull{cnn}s, although achieving remarkable results in object detection and classification challenges, still are not robust enough for many applications. They are sensitive to ambient light conditions \cite{Ref45} and to additive noise \cite{Ref9} as a result of pockets in their manifold. These algorithms, are based on bottom-up object detection and recognition process. As such, they do not include the means to use top-down contextual information for a more holistic process. 

In the human brain, for comparison, there are quick projections of \acrfull{lsf} information, in parallel to the bottom-up systematic progression of the image details along the visual pathways. This coarse but rapid blurred representations are used for an initial rough classification hypothesis with low discrimination. These hypotheses are followed by higher discriminability tests for an overall improved efficiency of object recognition in cases of ambiguity. This model was proven by a series of \acrfull{fmri} and \acrfull{meg} studies, which showed early \acrfull{ofc} activation by \acrshort{lsf} information, preceding the corresponding object recognition activation in \acrfull{itc} by about 50 ms \cite{Ref35}.

\begin{figure}
  \centering
  \includegraphics[width=0.8\textwidth]{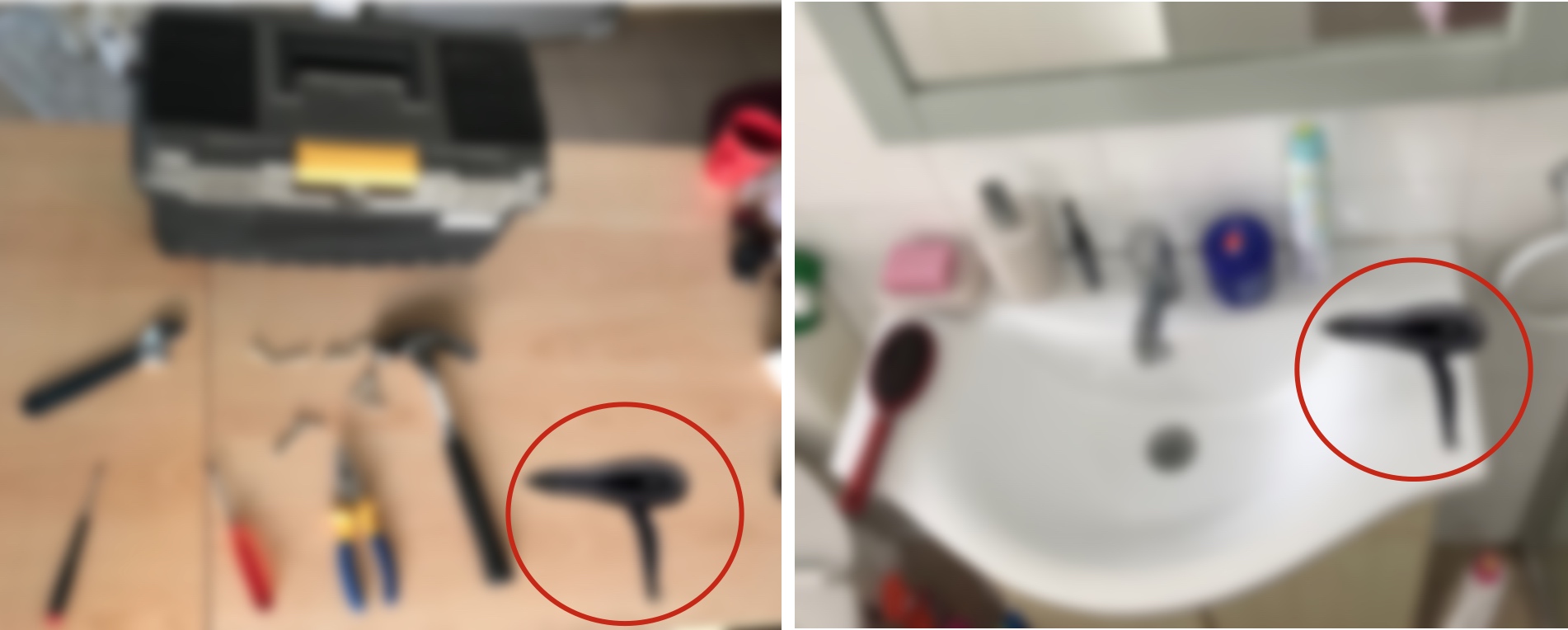}
\caption{The powerful effects of context.}
\label{fig:0}       
\end{figure}

The idea that context-based predictions make object recognition more efficient is prominent in Fig. \ref{fig:0}. The hairdryer in the right image and the drill in the left image are identical objects. Nevertheless, contextual information uniquely resolves ambiguity in each case. In this case, recognition cannot be accomplished quickly based only on the physical attributes of the target; contextual information provides more relevant input for the recognition of that object than can its intrinsic properties.

The working assumption underlying this research is that setting the human visual system as a source of inspiration and getting better in imitating its architecture and mechanisms, will enable us to develop a more robust and accurate object recognition algorithm with better results than exists today. Specifically, integrating the state of the art bottom-up feed-forward algorithms (\acrshort{cnn}s), together with the top-down mechanisms of the human visual system. The main contributions in this paper can be highlighted in the following aspects:
\begin{enumerate} 
\item Design of a biologically inspired architecture for integration of top-down and bottom-up processes into one holistic solution (section \ref{proposal}).
\item Demonstration of the importance of contextual-based predictions and their potential to improve the performance of bottom-up computer vision algorithms, such as deep \acrshort{cnn}s (section \ref{sec:10}).
\item Evaluation of the performance of a methodology for continuously updating models based on encounters with visual stimuli different than the ones they have been trained on (section \ref{sec:11}).
\end{enumerate}

The rest of the paper is organized as follows: Review and analysis of the different pathways and mechanism of the human brain, focusing on the "associative generation of predictions" mechanism, in section \ref{brain}. Review of \acrshort{cnn}s and their vulnerability to perturbations from the dataset they were trained on, in section \ref{cnn}. Finally, discussion and future work are summarized in section \ref{discussion}, followed by concluding remarks in section \ref{conclusions}.

\section{The Human Visual System}
\label{brain}
It is well known that the feature hierarchies in neural networks are often compared with the structure of the primate visual system, which has been shown to use a hierarchy of features of increasing complexity, from simple local features in the primary visual cortex to complex shapes and object views in higher cortical areas. Until recent years, the leading analysis of the object recognition process in the human cortex was focused on the ventral visual pathway that runs from the primary visual cortex, V1, over extra-striate visual areas, V2 and V4, to the inferotemporal cortex \cite{V1V4}. While most neurons show specificity for a certain object view or lighting condition \cite{Ref11}, the \acrshort{itc} cells show robustness in their firing to stimulus transformations such as scale, position changes, and light conditions. In addition, the \acrshort{itc} has been seen as a major source of input to the \acrfull{pfc}, which is seen as "the center of cognitive control" involved in linking perception to memory. This traditional bottom-up view of the human visual system strongly resembles the nature of neural networks, showing strong specificity in the first layers and more robust and complex activations in the higher layers. Nevertheless, this bottom-up view arises from the fact that most of the early research activities were focused on the low-level parts of the human visual system, due to the complexity involved in analyzing the higher-level parts of the human visual system. Extensive research in recent years that used advanced techniques, such as accurate \acrshort{fmri}, shows the main role of the high-level parts of the visual system, which integrate top-down signals with the bottom-up signals of the ventral pathway.

\begin{figure}
  \centering
  \includegraphics[width=0.6\textwidth]{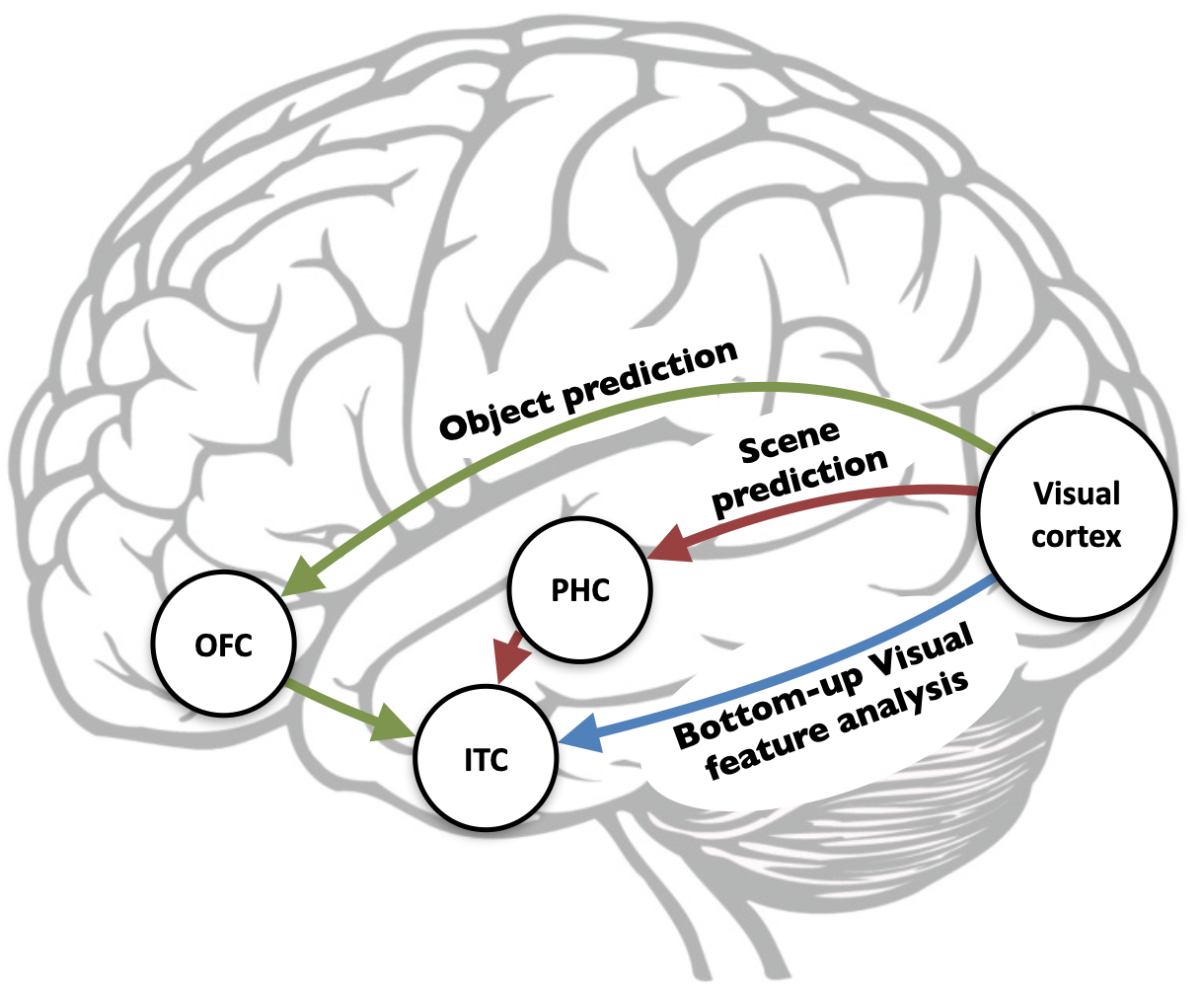}
\caption{Schematic description of pathways in the human visual system.}
\label{fig:1}       
\end{figure}

In general, the human visual system has two main pathways that participate in the continuous processing of the visual input. The ventral stream, which is a bottom-up feedforward visual features analysis that plays a major role in the perceptual identification of objects, and the dorsal stream, which governs the analysis of objects' spatial location and mediates the required sensorimotor transformations \cite{Ref12}. In addition to these pathways, there are two additional top-down projections: (1) The object prediction projection, which brings the low spatial information of the object to the orbitofrontal cortex that in turn uses this information to create fast gross object recognition; and (2) the scene prediction projection, which brings the low spatial information of the entire field of view to the parahippocampal cortex that in turn uses this information for scene recognition and contextual hypothesis \cite{Ref13,Ref14} These two top-down projections merge with the bottom-up ventral pathway in the inferotemporal cortex, as illustrated in Fig. \ref{fig:1}. The inferotemporal cortex integrates the three inputs and uses the top-down hypotheses to bias the recognition process.

Furthermore, latest experiments on the role of vision in motor control suggest that predictions are also key elements in the visual learning process. The difference between the brain's predictions and the received visual feedback is used as a trigger for the learning process.
when the desired goal is extrinsic, such as reaching or grasping, the brain processes the visual feedback in order to analyze the intended outcome (the target) and the actual outcome (the position of our limb relative to the target) in order to modify our behavior. Studies that have used \acrfull{eeg} have found that visual observation of a movement error elicits an \acrfull{erp} component with a timing and topography consistent with the \acrfull{frn}\cite{Ref15}. When the goal is intrinsic, e.g., the objective is sensation, there is no pre-defined goal- state, but rather there is an intrinsic motivation to learn about the environment in the most efficient manner, also known as active sensing \cite{Ref16}. Both cases of reinforcement-driven learning, whether intrinsic or extrinsic, support a framework, which suggest that humans and animals perceive the world by attempting to predict their sensations. Errors in prediction result in novelty that triggers learning and activates action for further exploration.

This paper suggests that an artificial agent operating in the world should use both its own predictions of the world and the perceived visual feedback for a reinforcement-learning mechanism, in order to improve the detection and recognition model of the platform. State of the art results with deep neural networks are largely dependent on access to large labeled datasets relevant for the attended task. In narrow and unique domains for which large labeled datasets don't currently exist and the initial data set is small, ongoing visual reinforcement learning can enable the possibility of exploiting the unsupervised stream of visual data that the unmanned agent receives during its operation to improve its recognition abilities with time. This could also potentially drive even better performing computer vision neural networks for conventional tasks, since they will be benefiting from models trained on gradually increasing datasets.

In addition, in 2015, researchers used multi-electrode arrays to measure hundreds of neurons in the visual ventral stream of nonhuman primates and compared their firing to the object recognition performance of more than 100 human observers. The results show that simple learned weighted sums of firing rates of neurons in monkey \acrshort{itc} accurately predicted human performance \cite{Ref6}. As we are aiming to design a biologically inspired architecture, this means that simulating the operation of the IT is a key factor in successfully creating an architecture that has a human-like accuracy and robustness.

Attempts to combine bottom-up and top-down processes in a computational model have been made in the past, by researchers from the computer vision field and the brain research field. Nevertheless, due to different research goals and technical barriers, these attempts were mainly single field centric. While the goal of brain researchers was mainly to build a model that will predict eyes' behavior and neural activity \cite{Ref17,Ref18,Ref19,Ref20}, computer vision researchers' main goal was to improve overall performance on specific tasks (segmentation \cite{Ref21}/detection \cite{Ref22}/tracking \cite{Ref23}/scene understanding \cite{Ref24}) with little or no bio-inspired architecture. In addition, there was a technological gap to design, train, and use a CNN-based system with the complexity of multi-level lateral and recurrent connections that exist in the human visual system.

\section{Convolutional Neural Networks}
\label{cnn}
In recent decades, the human brain inspired researchers to develop Convolutional Neural Networks. One of the main benefits of \acrshort{cnn}s are their inherent ability to use labeled datasets to automatically train layers of convolution filters to create task-specific feature space. Until recently, \acrshort{cnn}s did not manage to keep up with the state of the art traditional computer vision algorithms. However, in the last few years, the publication of large image databases such as ImageNet, the advancement of high performance computing systems, and the improvements in training deep models, boosted \acrshort{cnn}s' performance and enabled them to gain popularity and achieve dramatic progress on object classification tasks \cite{Ref1,Ref2,Ref3}. In addition, empirical evidence \cite{Ref4} as well as mathematical proofs \cite{Ref6,Ref5,Ref7} show that deep \acrshort{cnn}s, with tens and hundreds of layers, have a high expressive power and can create complex feature space, which outperforms traditional computer vision algorithms, as well as shallower \acrshort{cnn}s.

Generally speaking, the output layer unit of a neural network is a highly nonlinear function of its input. It represents a conditional distribution of the label given the input and the training set. It is therefore possible for the output unit to assign non-significant probabilities to regions of the input space that contain no training examples in their vicinity. Such regions can represent, for instance, examples of the same objects the \acrshort{cnn} was trained on, but from different viewpoints. These examples, though belongs to the same objects and share both the label and the statistical structure of the original inputs, can be relatively far in pixel space and be misclassified.

Even more, we expect imperceptibly tiny perturbations of a given image not to change the class at the output and that an input image that is in the vicinity of a given training input x (in pixel space) to be assigned a high probability of the correct class by the model. Nevertheless, it has been shown that for deep neural networks, this smoothness assumption does not hold. Specifically, one can find adversarial examples, which are obtained by imperceptibly small perturbations to a correctly classified input image, so that it is no longer classified correctly \cite{Ref8}.
Adversarial examples represent low-probability (high-dimensional) "pockets" in the manifold \cite{adversial}. These perturbed examples are called "adversarial examples" and have been shown to exist in neural networks with varied numbers of layers and varied training data. Although the common practice in training \acrshort{cnn}s includes employing input deformations to the training dataset in order to increase the robustness of the models, it doesn't solve the "blind spots" in the manifold of the trained neural network. This phenomenon enables easily fooling the neural network as demonstrated recently in \cite{Ref9} and \cite{Ref10}.

In addition, recent successes of deep reinforcement learning (RL) techniques have raised widespread interest in their potential for solving practical problems, with most famous cases being in the field of autonomous driving \cite{deeprldrive} and game playing \cite{deeprlgame1, deeprlgame2}. However, most state of the art algorithms for RL require vast amounts of data and training time. Decomposing the problem into state inference, value updating and action selection components, as is done in the brain, may allow for more efficient learning and the ability to track changes in the environment on fast timescales, similar to biological systems \cite{deeprl}.

\section{The proposed architecture}
\label{proposal}
\begin{figure}[hb]
  \centering
  \includegraphics[width=0.6\textwidth]{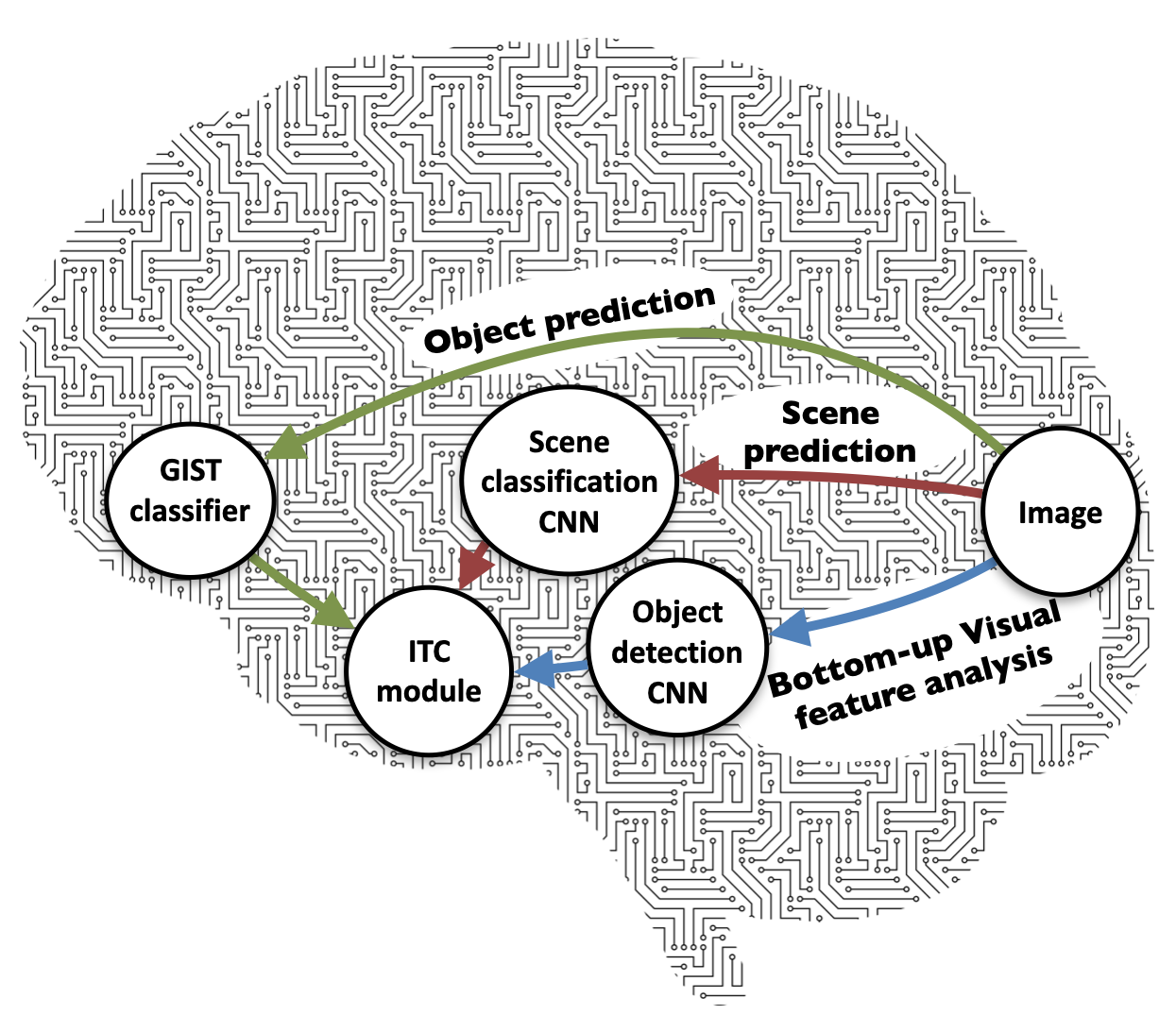}
\caption{Top-level schematic illustration of the pathways in the VAP model.}
\label{fig:2}       
\end{figure}

Our hypothesis is that the sensitivity of the neural networks comes from their bottom-up feed-forward nature and from their limited ability to include the time space in their analysis of the incoming video stream. The proposed Visual Associative Predictive (VAP) architecture is based on the concepts behind the integration of the top-down and bottom-up processes of the human visual system, combined with state of the art object recognition models, e.g., deep \acrshort{cnn}s. As a first step, the proposed model was designed to have a similar structure to that of the human visual system. The pathways in the VAP model, which imitates those illustrated in Fig. \ref{fig:1}, can be seen in Fig. \ref{fig:2}. We show that imitating these top-down mechanisms, together with the state of the art bottom-up feed-forward mechanism, creates a target recognition model, which is more accurate and enjoys better robustness than using a bottom-up deep \acrshort{cnn} model alone.

The proposed model also includes implementation of the human brain visual feedback driven learning mechanism. This learning mechanism is based on imitating two layers of the human visual system. The first layer, an Object-file mechanism, is a midlevel visual representation that "sticks" to a moving object over time, stores and updates information about the object's properties \cite{Ref25}. In the second layer, the reinforcement learning system uses the Object-file information for autonomous refinement of the recognition model. 

To generate the information required for its operation, the VAP model uses three types of classifiers. For the bottom-up object classification process we used the \acrfull{frcnn} \cite{rcnn} with the VGG16 model \cite{Ref3}. For the top-down scene prediction we used the Places205 deep \acrshort{cnn} \cite{Ref26}. Finally, for the top-down object prediction we used a "gist" classifier (described in \ref{sec:4}) that gives gross classification of the objects into two categories: man-made or natural. The outputs of the three classifiers are sent to a module that aims to imitate the operation of the \acrshort{itc} and integrate the three signals into one decision. 

\subsection{Attention mechanism}
\label{sec:3}
\begin{figure}
  \centering
  \includegraphics[width=0.9\textwidth]{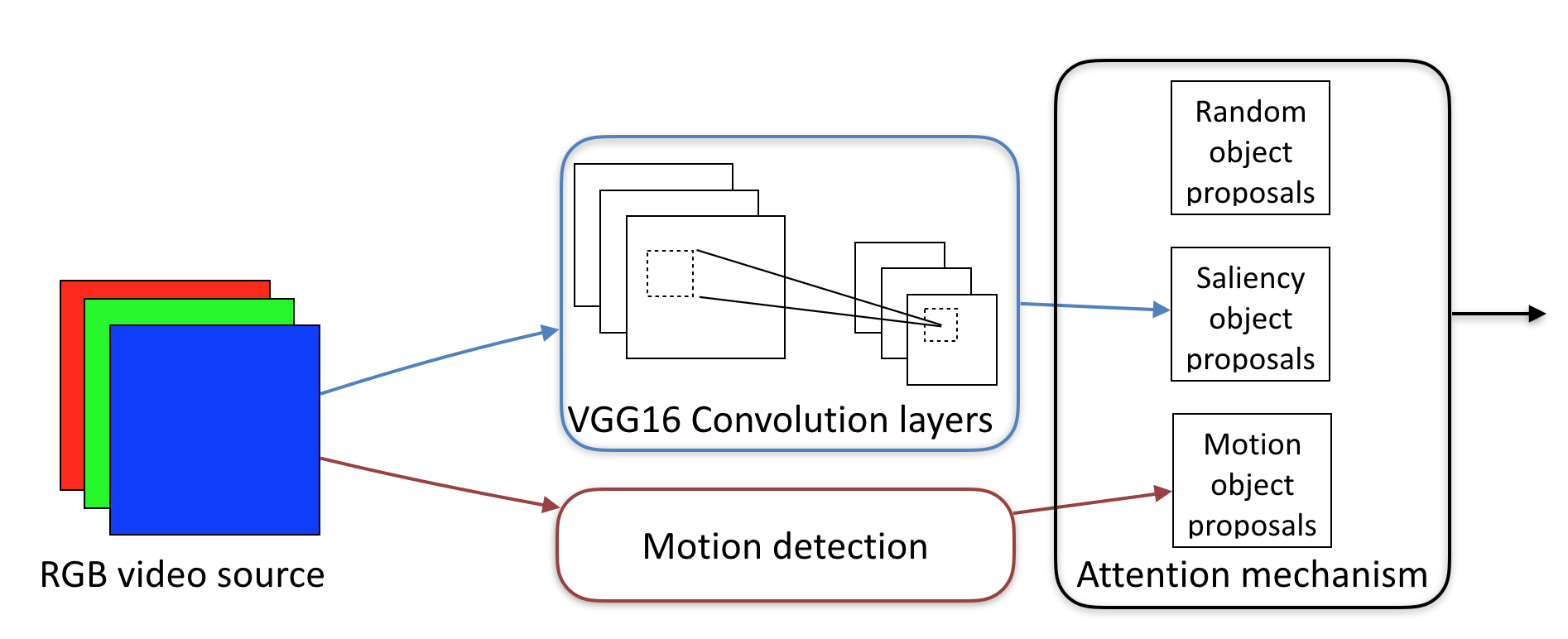}
\caption{Schematic illustration of the VAP model attention mechanism.}
\label{fig:3}       
\end{figure}

In order to improve its efficiency, the algorithm does not process the entire image at once, but rather processes a bounded number of small areas, using the attention mechanism illustrated in Fig. \ref{fig:3}. The attention of the algorithm is determined according to three biologically inspired attention generators. The first mechanism is the saliency detection mechanism. Saliency stands for the quality by which an object stands out relative to its neighborhood. Our eyes do not move smoothly across the scene, instead they make short and rapid movements between areas containing more meaningful features. Saliency detection is considered to be a key attentional mechanism that facilitates learning and survival by enabling organisms to focus their limited perceptual and cognitive resources on the most pertinent subset of the available sensory data \cite{Ref27}. In our model, we have used the \acrshort{frcnn} natural structure and its \acrfull{rpn} module, which is based on the activity feature map created by the convolutional layers, as a source for saliency objects proposal.

A motion detection algorithm was used to draw attention to moving objects. Visual motion is known to be processed by neurons in the primary visual cortex that are sensitive to spatial orientation and speed and are considered a pre-attentive mechanism, which help focus attention on moving objects. A simple motion detection algorithm is used, which conducts two main operations: foreground-background separation and adjacent frames subtraction. The foreground-background separation is based on pixel value statistics \cite{Ref28}. The frames subtraction was performed on five adjacent frames in order to detect local changes caused by moving objects. The results of the two algorithms were merged into single object proposal mechanism.

Finally, a random attention generator was employed to focus attention on random locations in the scene, in order to complement the other two search mechanisms. This random generator imitates the saccades movement of the human eyes that make rapid random jittering movements while scanning a visual scene, even when we are fixated on one point. The random attention generator proposes bounding boxes for exploration that are between the saliency and the motion detector bounding boxes. The output of the attention mechanism is a list of bounding boxes object proposals that is passed forward for further processing, as can be seen in Fig. \ref{fig:6}.

\subsection{Object prediction}
\label{sec:4}
The gross classification of the objects in the VAP model are based on the observation made by Antonio and Aude \cite{Ref29} about the nature of the power spectrum difference between man-made and natural objects. They suggest that the differentiation among man-made objects resides mainly in the relationship between horizontal and vertical contours at different scales, while the spectral signatures of natural objects have a broader variation in spectral shapes. The object prediction module takes each image of an object and calculates its spatial envelope model, also known as the "gist" descriptor \cite{Ref30}, which is a low-dimensional representation that encapsulates the spatial structure of the object. The descriptor is based on Friedman work \cite{Ref31}, who suggested that an abstract representation of a scene, which he named "gist", spontaneously activates memory representations of scene categories (a city, a mountain, etc.)

The "gist" descriptor divides each object into 4x4 blocks. In each block, the "gist" features are calculated using three frequency scales 0.02, 0.08, and 0.32 cycles/pixel. For the first two scales, eight orientations are used and for the third scale, four orientations are used. In total, the "gist" descriptor generates 960 (= 3 colors * 16 blocks * [2*8 orientations + 1*4 orientations]) features. A State Vector machine was pre-trained on the 19,319 segmented objects from the SUN2012 dataset. The VAP model uses this pre-trained \acrfull{svm} to classify each object into man-made or natural object. For example, cars and bicycles will be classified as man-made objects, while people and cats will be classified as natural objects. This gross classification and its score are sent to the IT module to be integrated with the other VAP pathways.

\subsection{Scene contextual prediction}
\label{sec:5}
The scene contextual prediction module contains the Scene-Object (SO) matrix and the Object-Object (OO) matrix that represent the cumulative contextual information the robotic agent gathered during its operation. The size of the SO matrix SxO is determined by the number of scenes S the scene classifier can recognize and the number of objects O the bottom-up \acrshort{cnn} can recognize. The SO matrix easily encapsulates contextual probabilities about the dataset, such as that cows are seen mainly in "open country" and scenes, while seagulls are seen mainly in "coastal" scenes. The OO matrix in turn encapsulates the model's information on objects that tend to co-appear. 

During the entire operation of the robotic platform, the VAP model uses the Places205 deep \acrshort{cnn} for scene classification. Once the current scene classification is determined, it is accompanied by information on the typical sets of objects that tend to co-appear within that context. This joined information sets the model's expectation in the current context. The VAP model then uses the global scene classification as well as the contextual cues that are provided by the objects in the scene that have been detected so far, to generate a vector of probabilities containing the probability for each object class to be seen in the current scene. This vector of probabilities is sent to the \acrshort{itc} module to be integrated with the other VAP pathways. In parallel, the continuous stream of information is used to update the scene contextual prediction module. In this way, the SO matrix is kept dynamic and the agent's knowledge about the world get richer and better the more it operates in the world.

\subsection{Inferotemporal cortex module}
\label{sec:6}
The problem of combining top-down and bottom-up cues is currently not solved in computer vision. As a result, while it is accepted that as part of the recognition process, the brain performs seamless fusion of pre-attentive automatic bottom-up processing and attentive task-selective top-down processing, current object recognition models, such as deep \acrshort{cnn}, still focus on a bottom-up process. The proposed VAP model suggests a method for the rapid feed-forward process to engage the top-down contextual information in order to improve the classification in cases of ambiguity.

In the human visual system the bottom-up \acrfull{hsf} ventral stream meets the top-down \acrshort{lsf} scene and object hypothesis in the inferior temporal cortex. Similarly, in the VAP model the bottom-up stream and the top-down prediction meet in a module named the inferior temporal cortex mechanism. It is important to note that there is evidence to suggest that the integration of top-down and bottom-up processes in the human brain is not exclusive to \acrshort{itc} and may also take place in other areas of the brain. Nevertheless, we limited our scope of work to the major role of the \acrshort{itc} in this process.

The VAP model decision is based on adding a correction factor to each element in the probabilities vector given by the bottom-up classification process. The \acrshort{itc} mechanism in the VAP model uses two functions to regulate the integration of the two top-down pathways into the bottom-up pathway. First, a Gaussian function is used to regulate the correction factors magnitude and give a higher correction factor in cases of ambiguity. The Gaussian function, which has high values in the center and low values at its edges, expresses the notion that if the bottom-up classifier is "pretty sure" that the analyzed object is very similar (probability above a specific threshold) or very different (probability below a specific threshold) from the object model stored in the long term memory module, the correction should be minimal. This correction factor magnitude regulator, R\textsubscript{Magnitude}, is given by:
\begin{equation}
R_{Magnitude} = ae^{-\frac{\left(P_{BottomUp}-\mu\right)^2}{2\sigma^2}}
\end{equation}

Where, P\textsubscript{Bottom-up} is the probability vector from the convolutional neural network. $\mu$ and $\sigma$ are the parameters governing the shape of the R\textsubscript{Magnitude} regulation function and where set to $\frac{1}{5}$ and $\frac{1}{3}$, respectively. Second, a hyperbolic tangent function (tanh) is used to regulate the sign of the correction factors. The hyperbolic tangent function, which has positive values on one side and negative values on the other side, expresses the notion that for objects with high probability to appear in the current context the correction will be positive and that for objects with low probability to appear in the current context the correction will be negative. This correction factor sign regulator, R\textsubscript{Sign}, is given by:
\begin{equation}
R_{Sign} = tanh\left(b\left(P_{Object|Context}-c\right)\right)
\end{equation}

\begin{figure}
  \centering
  \includegraphics[width=0.9\textwidth]{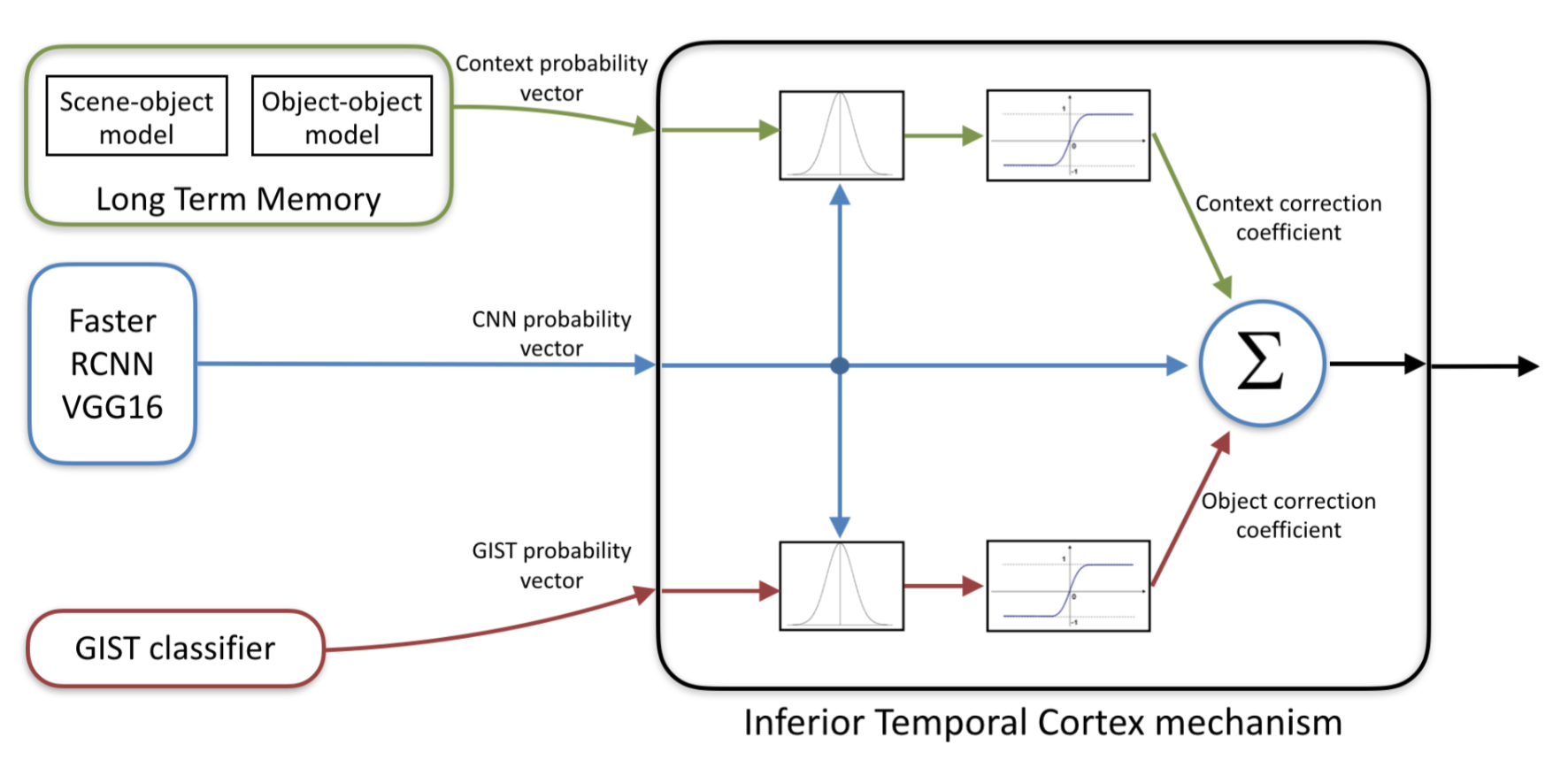}
\caption{Illustration of the inferior temporal cortex mechanism in the VAP model.}
\label{fig:4}       
\end{figure}

Where, P\textsubscript{Object$\mid$Context} is the probability of each object to be in the scene given the current context. $b$ and $c$ are the parameters governing the shape of the R\textsubscript{Sign} regulation function and where set to $10$ and $\frac{1}{8}$, respectively. Following this logic, the correction coefficients for each top-down pathway is calculated as follows:
\begin{equation}
C_{Context} = P_{Context}*ae^{-\frac{\left(P_{BottomUp}-\mu\right)^2}{2\sigma^2}}*tanh\left(b\left(P_{Object|Context}-c\right)\right)
\end{equation}

where P\textsubscript{Context} is the confidence level in the context classification. Figure \ref{fig:4} illustrates this structure of the ITC mechanism. 

For a specific scene, the scene-dependent prior probability is extracted from the SO matrix. For the set of objects currently in the scene, the OO matrix is used to extract the co-occurrence probability and use it as an addition to the prior probability factor. The SO and OO matrices are being continuously updated according to the information gathered by the VAP model during its operation. Therefore, more accurate and continuously improving context-dependent prior probabilities are being generated with time. As a result, the VAP model is more accurate and more robust as the prior probabilities become more accurate.

\subsection{Object files}
\label{sec:7}
As mentioned, it is widely accepted that the brain attaches an Object-file to each moving object in its field of view. The Object-file stores and updates information about the changes in the object's appearance and properties with time and plays a critical role in the retrieval of the object's previous characteristics, some of which may no longer be visible \cite{objectfile}. We implemented the object-file mechanism in the following way. The object-file mechanism includes a Kalman filter-based tracker that tracks each moving object in the scene from the moment it enters until it is out of the field of view. During the motion of the object in the field of view, the object-file attached to it stores the output of the classification process in each frame. It then filters out observations with high Mahalanobis distance for all classes, which may result in false classification.

The Object-file is integrated in the classification process of moving objects in three ways. First, in cases where the initial observations of an object are all with low probabilities for all classes, the Object-file mechanism will pick those with the highest probabilities. When samples with higher probabilities will begin to arrive, it will filter out the samples it used so far and will base its decision only on the top observations. In this way, the VAP model imitates the object-file mechanism in the human visual system, gradually increasing its confidence in the classification of an object, the longer the object stays in the field of view. In addition, the Object-file mechanism uses only samples whose Mahalanobis distance to at least one class is low enough. When some aspects of the object are no longer visible and the classification probability drops, it uses information from previous frames that resulted in high probability score. As such, the Object-file mechanism enables the VAP model to retain perceptual continuity, even in periods when the object's classification is ambiguous due to challenging views, such as partial occlusion and shade. A final contribution of the Object-file mechanism is that it gathers information on the object that is later used by the reinforcement module in order to improve the classification process.

\subsection{Reinforcement learning module}
\label{sec:8}
Adding reinforcement learning abilities to the robotic agent reduces the burden of the information gathering for the training phase, while enabling it to improve its classifier using the visual data encountered during its operation. For example, in the experiments done using the VAP model, the training was done on an open source dataset. This dataset, like others, contains mainly images taken by a photographer that was located on the same surface as the object. However, the experiments of the VAP model were conducted using an elevated camera with a top view of the objects. Therefore, the initial classification performance of the VAP model was lower than expected. Nevertheless, as the reinforcement module accumulated information on the objects in the scene, the VAP model became more accurate.

\begin{figure}
  \centering
  \includegraphics[width=0.99\textwidth]{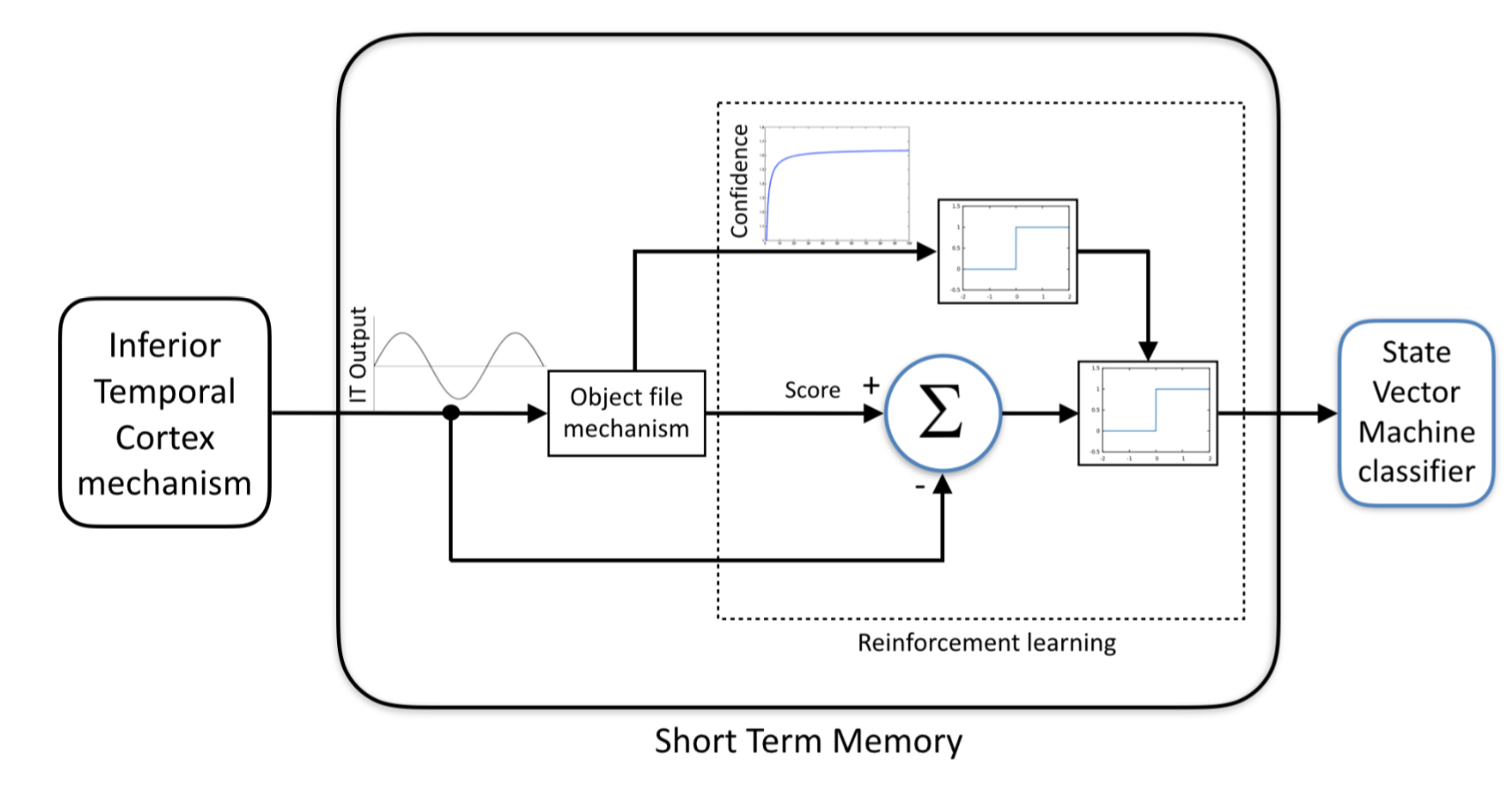}
\caption{Illustration of the reinforcement mechanism in the VAP model.}
\label{fig:5}       
\end{figure}

An object moving around the scene may be varied visually in a number of ways, due to the new viewing conditions in each frame. For the sake of terminology, each specific realization variation of an object, which may result in a different feature vector, is called an instance. For each object tracked by the object-file mechanism, the reinforcement learning mechanism is looking for instances where there is a gap between the instantaneous visual feedback received from the \acrshort{cnn} in the current frame and the stable object-file semantic description of the objects. Meaning, after the Object-file has accumulated enough instances to reach high confidence level recognition of the object, the reinforcement learning mechanism compares all new incoming instances from the \acrshort{itc} mechanism to this representation of the object, as can be seen in Fig. \ref{fig:5}. When a new instance creates a false prediction, it evokes a learning process, which uses the features of this instance to refine the model.

The proposed model uses an \acrshort{svm} coupled with a \acrshort{cnn} as part of a multistage process. A \acrshort{cnn}, which is an efficient mechanism for learning an invariant features space from images, is used as a first stage. The deep \acrshort{cnn} is first trained to learn good invariant representations and then used on incoming frames as a feature extraction mechanism. As a second stage, the feature vectors are fed into a kernel \acrshort{svm}, which produces good decision surfaces by maximizing margins. It has been shown that this multistage process and its variations usually result in similar performance \cite{Ref32,Ref33,Ref34} to the fully connected layers of a \acrshort{cnn}.

Working in a high-dimensional feature space increases the generalization error of support vector machines, as there are not enough data points in the training set to accurately define the boundaries between the classes. Our reinforcement learning model detects and stores samples in video sources that: (a) belong to a specific class but were on the wrong side of the hyper plane and were wrongly classified; or (b) close to the hyper plane where there are naturally fewer samples as it is far away from the mean of the distribution of the class. It then uses these samples to refine the \acrshort{svm} decision function when there is no more movement in the scene.

\section{Results}
\begin{figure}
  \centering
  \includegraphics[width=0.99\textwidth]{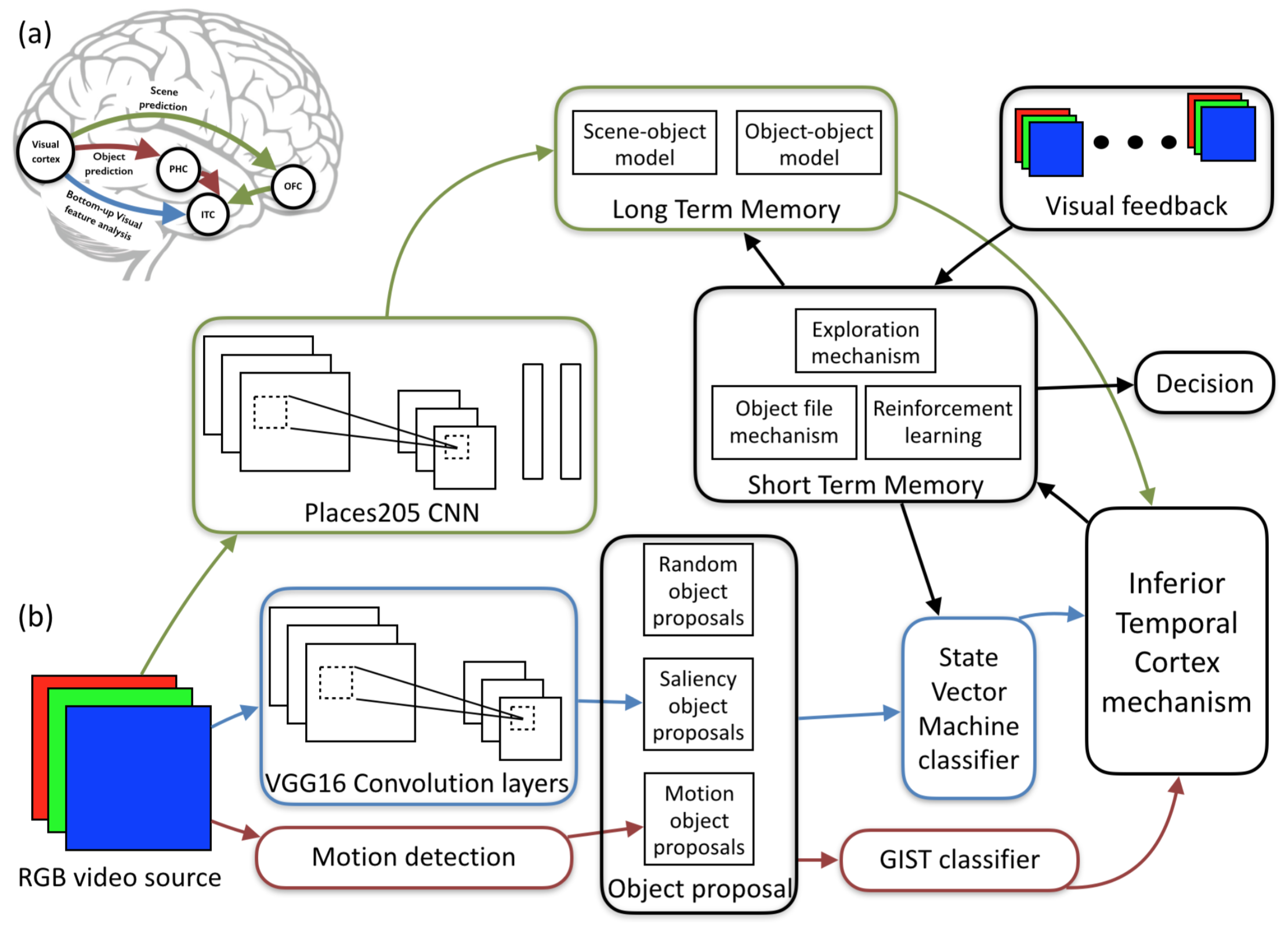}
\caption{Schematic illustration of (a) the main pathways in the human visual system and (b) the pathways in the VAP model, using the same color code.}
\label{fig:6}       
\end{figure}

In this work, a novel architecture for object recognition in video sources is proposed. The Visual Associative Predictive (VAP) architecture integrated several biological mechanisms into one system that imitates the concepts behind the efficiency of the human visual system. Figure \ref{fig:6} demonstrates the similarity between the VAP model and the human visual system. Using the same color code (green for scene prediction pathway, red for object prediction pathway, and blue for the bottom-up pathway), this figure shows the pathways in the VAP model that correspond to those in the human visual system.

The SUN2012 dataset was used for the purpose of pre-training the VAP model. This dataset was chosen because the scene of each image in the dataset was labeled and all the main objects in each image were segmented and labeled as well. Overall, the SUN2012 covers 313,884 segmented objects from 908 scene categories. Nevertheless, a subset of the object categories was chosen, which had dozens of samples in the dataset and whose appearance was shown on numerous scenes. The chosen object categories are: cabinet, chair, person, bed, car, plant, plant pot, desk, sink, clock, sofa, bookcase, television, telephone, boat, shoe, washing machine, traffic lights, bicycle, teddy bear, cow, dog, crosswalk, conference table, and aircraft. In total, we got 19,319 segmented objects divided to 25 object categories that appear in 727 scene categories. In many use cases, it is not probable to have thousands of samples to train the classifier, not to say millions in the case of deep \acrshort{cnn}. In addition, the goal is for the model to improve with time the same way humans usually get a small set of examples from parents/teachers and enrich their classification model with time, by themselves. Therefore, only 2063 segmented objects were randomly chosen for training the model.

\subsection{Results on Images}
\label{sec:10}
First, the contribution of the contextual scene information on the recognition of objects in still images was examined. The 17,256 images from the SUN2013 dataset that weren't used for training, were classified using both the \acrshort{frcnn} with the VGG16 model and the VAP model. As explained, the fully connected layers of the \acrshort{frcnn} were replaced with \acrshort{svm}. Standardization was applied to the data fed to the \acrshort{svm}. The main advantage of standardizing is to avoid attributes in greater numeric ranges dominating those in smaller numeric ranges. Of course, the same standardization was used to scale both training and testing data.

\begin{figure}
  \centering
  \includegraphics[width=0.8\textwidth]{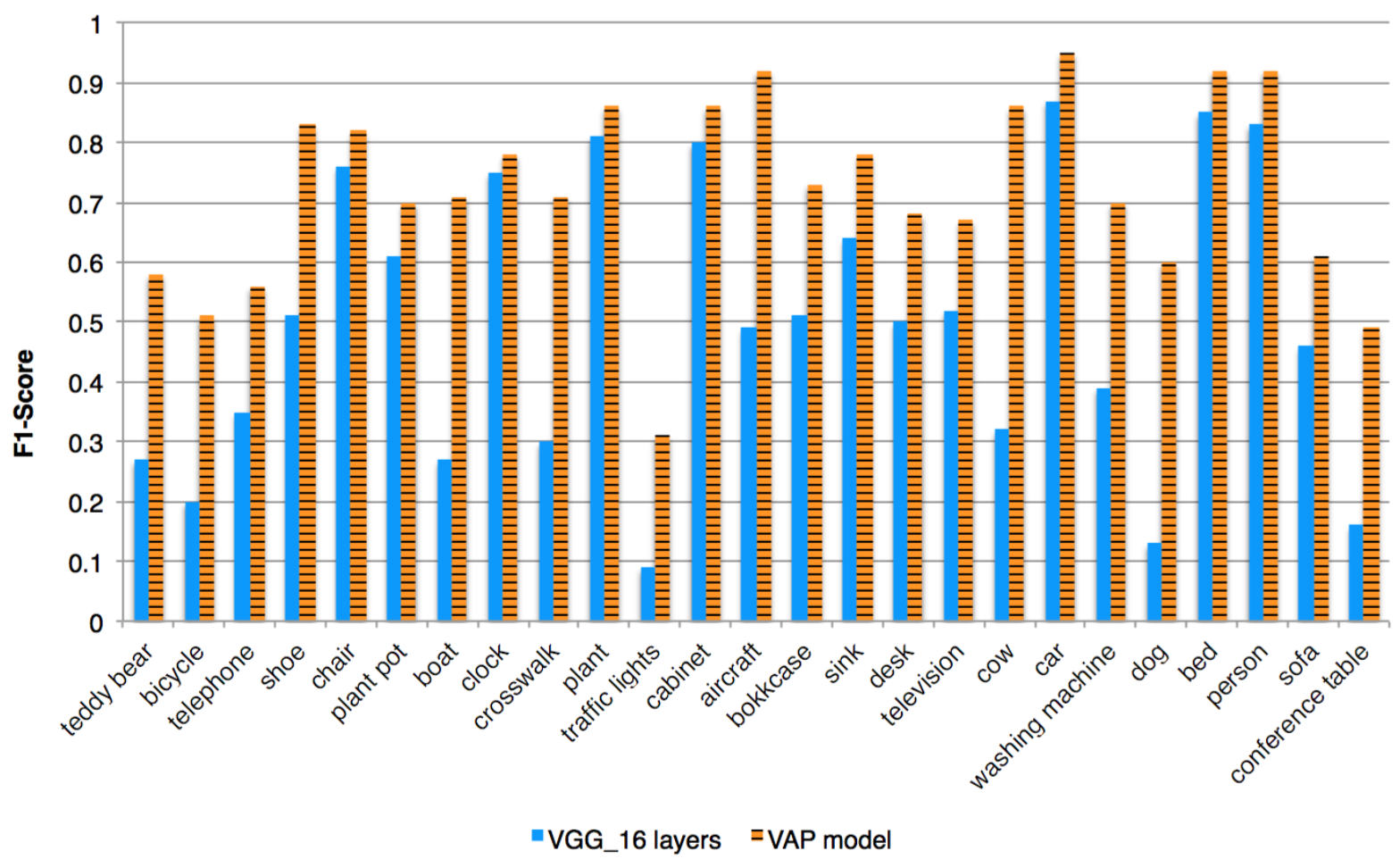}
\caption{Comparison for all classes using the complete VAP model and using only the bottom-up pathway (VGG16).}
\label{fig:7}       
\end{figure}

\begin{figure}
  \centering
  \includegraphics[width=0.8\textwidth]{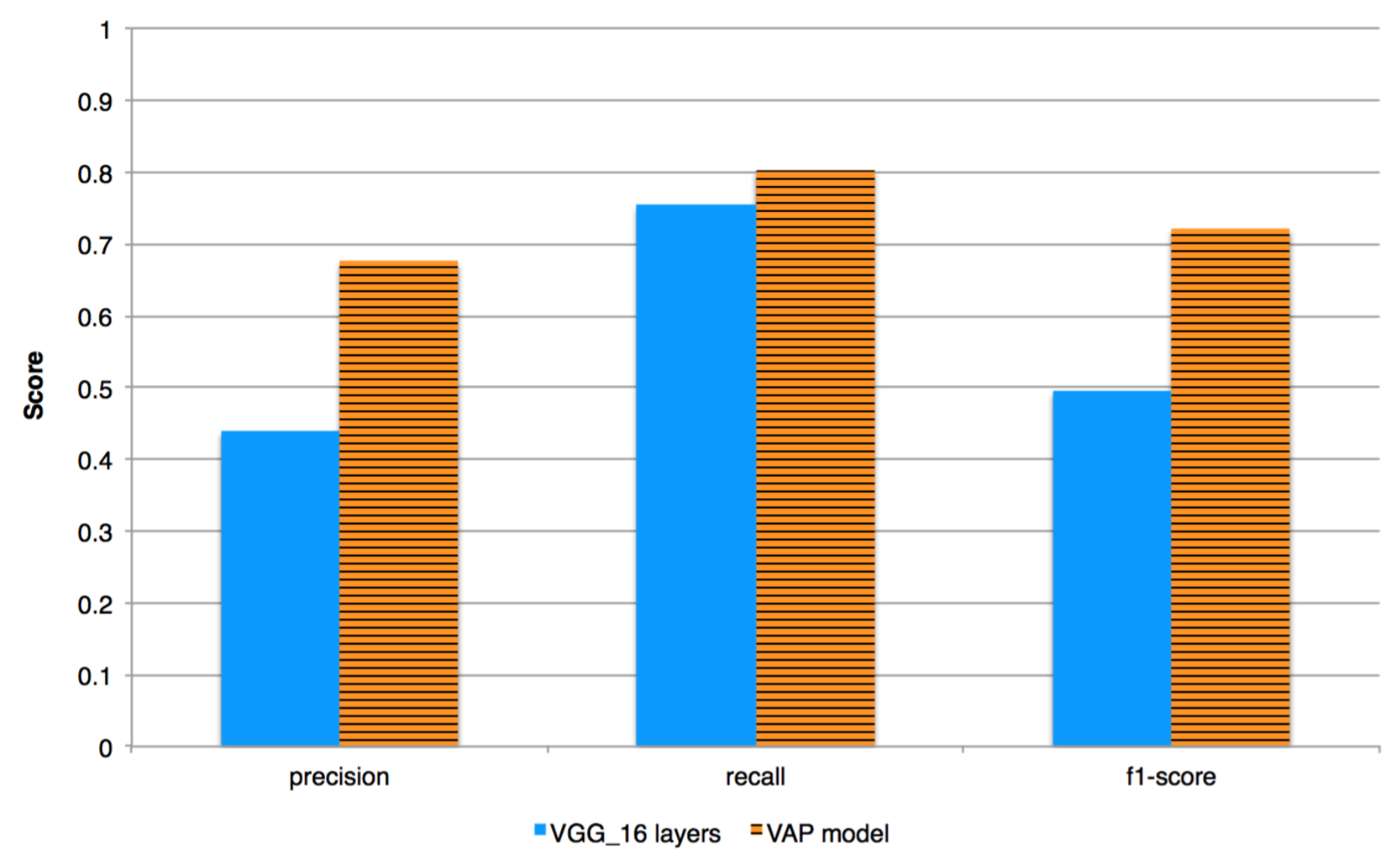}
\caption{Overall comparison between using the complete VAP model and using only the bottom-up pathway (VGG16). Precision, recall, and F1-score are shown.}
\label{fig:8}       
\end{figure}

The results, seen in Fig. \ref{fig:7}, show that the contribution of the contextual information resulted in better performance of the VAP model for all class categories. The F1-score, which is the harmonic mean of precision and recall, was used for the comparison, as it considered both the precision and the recall of the experiment results. The improvement is in all three measures (precision, recall, and F1-score), as seen in Fig. \ref{fig:8}.

\subsection{Results on Videos}
\label{sec:11}
For the evaluation of the proposed model on video streams, we used 39 short video clips filmed from approximately the same location. The location was a window on the third floor of a building located 50 meters from the filmed area. That position was used as it resulted in view angle different from the training set, which is usually from almost a zero angle. That way, we could observe the learning capabilities of the VAP model, which needed to cope with objects somewhat different than it was trained on and use its reinforcement mechanism to improve with time.

The video clips were taken over a period of a month in order to capture the scene in different lighting conditions and object arrangements. Each video clip was 45-150 seconds long, which resulted in a 30-minute long test set. The dataset includes 45,613 instances of 181 moving objects, mainly belonging to the categories: Car, Person, Bicycle, and Cat. We tested the proposed model on the video clips and evaluated the contribution of the contextual information, the Object-file mechanism, and the reinforcement-learning scheme, to its overall performance. For future work, the dataset is available at: \href{https://drive.google.com/drive/folders/0B_Tni066P8G2U3ZEN290QVBkbEU?usp=sharing}{VAP's video dataset}.

\begin{figure}[b]
  \centering
  \includegraphics[width=0.85\textwidth]{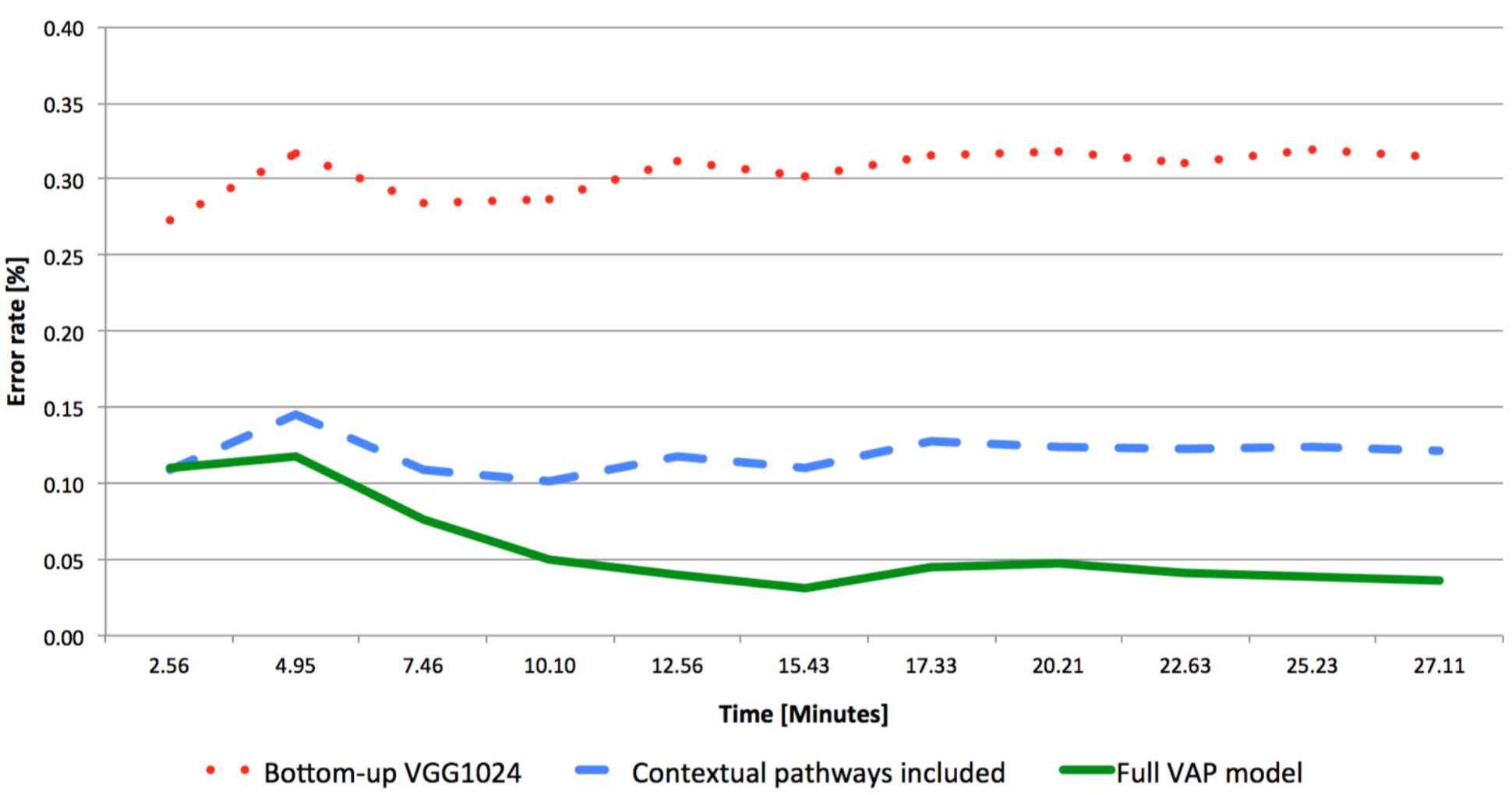}
\caption{Performance on video sources of the VAP model with the VGG1024 faster RCNN as the bottom-up classifier.}
\label{fig:9}       
\end{figure}

For the evaluation of the classifier's performance, the error rate was calculated by dividing the number of instances with false classification by the total number of instances. In Fig. \ref{fig:9} we can see the superior performance of the proposed model on the movie clips dataset. This figure shows the cumulative error rate, where the error rate is measured as the sum of false classified instances divided by the total number of instances. It is evident that the proposed model is learning from past experience, improves its accuracy, and outperforms the bottom-up \acrshort{frcnn} classifier (with VGG1024 model). Furthermore, even without applying the reinforcement learning, the model uses the information it has about the world, from the SUN2012 dataset, and uses this prior information to perform better than the bottom-up process alone.

\begin{figure}
  \centering
  \includegraphics[width=0.85\textwidth]{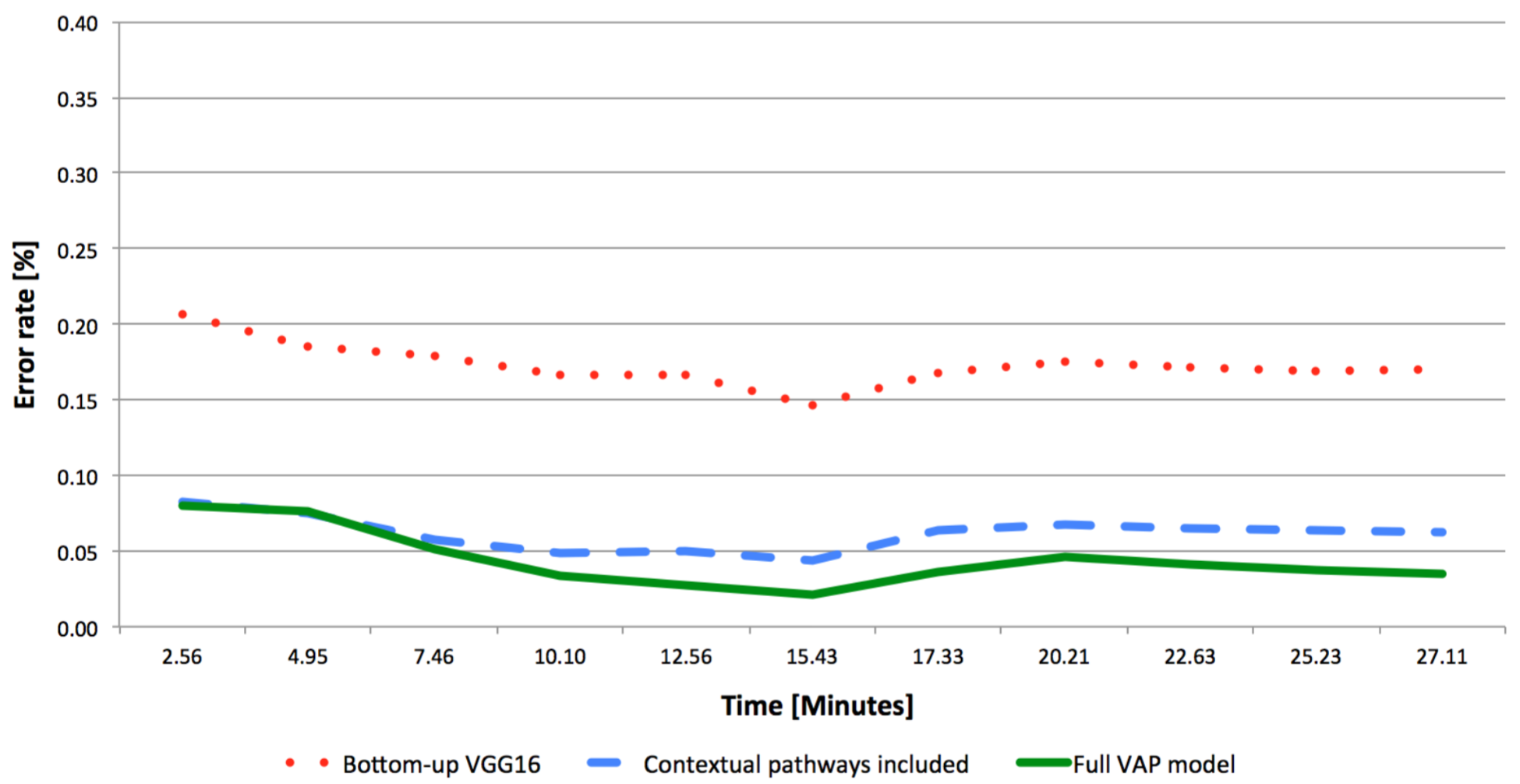}
\caption{Performance on video sources of the VAP model with the VGG16 faster RCNN as the bottom-up classifier.}
\label{fig:10}       
\end{figure}

These results emphasize the importance of the Object-file module for video sources as a tool to retain perceptual continuity, as well as to enable the reinforcement learning. Comparing the results between Fig. \ref{fig:9}, with the shallower VGG1024, and Fig. \ref{fig:10}, with the deeper VGG16, suggests that the superior performance of the proposed model is especially prominent for shallow \acrshort{cnn}s with reduced performance. When using shallow \acrshort{cnn}s, there are more cases of ambiguity where the ability of the Scene-Object-Matrix to contribute differentiating information is higher. In addition, the rate of false prediction is higher in theses cases and the reinforcement learning gains new information more rapidly. As a result, the proposed model converges faster and becomes extremely beneficial in cases of platforms with strict low computational resources that dictates shallow \acrshort{cnn}s, such as small autonomous systems.

\begin{figure}[hb]
  \centering
  \includegraphics[width=0.86\textwidth]{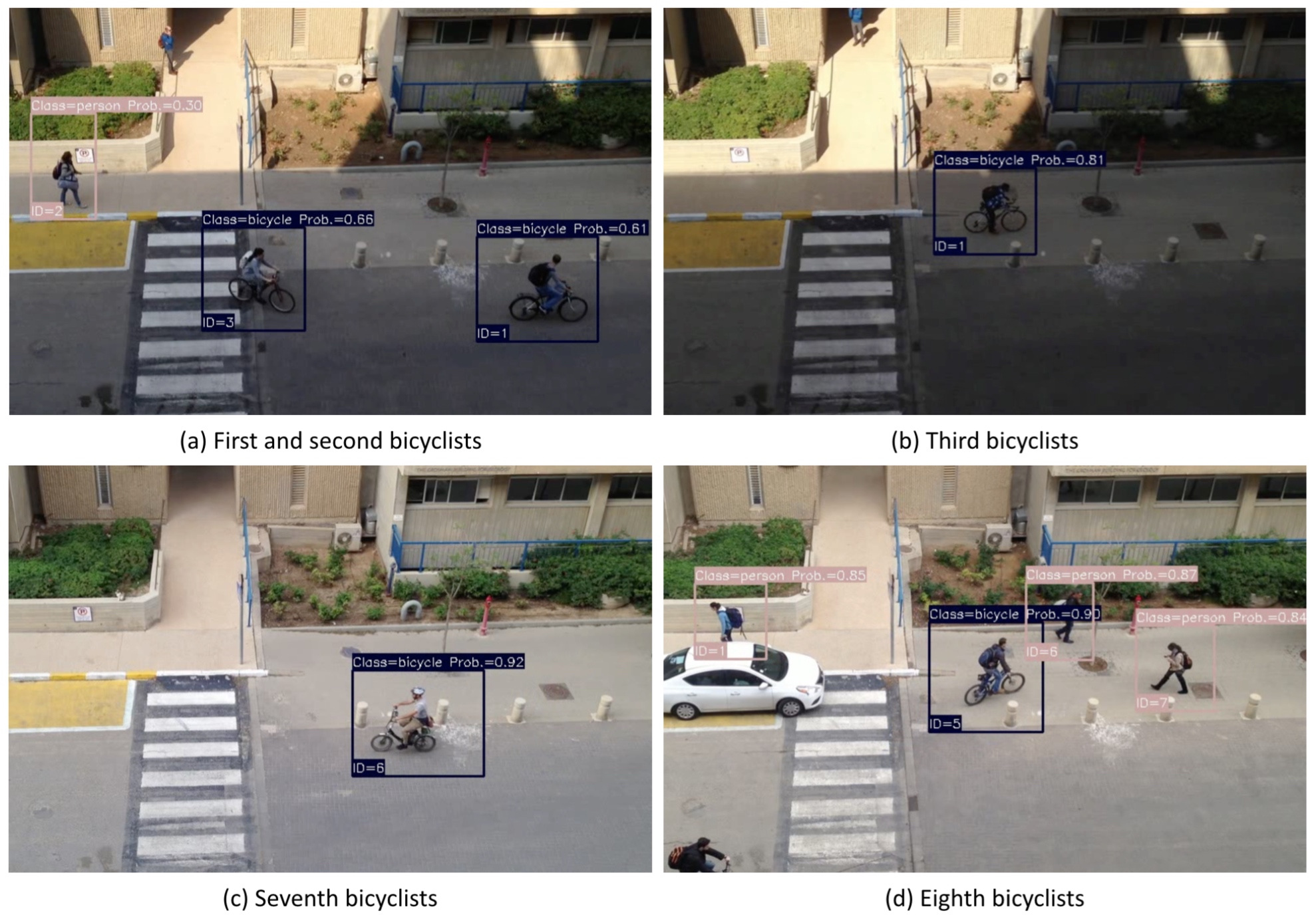}
\caption{Classification results of the VAP model for bicyclists passing through the test scene. Examining the confidence level of the system for (a) the first and second bicyclists, (b) the third bicyclist, (c) the seventh bicyclist, and (d) the eight bicyclist, shows gradual increase in the performance of the system.}
\label{fig:11}       
\end{figure}

As explained, the training dataset of SUN2012 includes mainly objects with low altitude angle view. As a result, when the system started to work it generated low scores for the objects picked up by the video motion detector. Nevertheless, after few minutes of watching the scene, it managed to learn from frames, which resulted in visual feedback that is different than the system prediction. These samples were used to update the \acrshort{svm}, by adding the upper view samples of the different classes to the existing low altitude angle training samples and refining the hyperplanes defining each class.
An example of the learning process can be seen in Fig. \ref{fig:11} that shows the classification and confidence level of the VAP model to the several appearances of bicyclists in the scene. It is evident that at the beginning the confidence level is low, but with time, the system's confidence increases as the reinforcement mechanism learns from past examples.

\newpage

\begin{figure}
  \centering
  \includegraphics[width=0.99\textwidth]{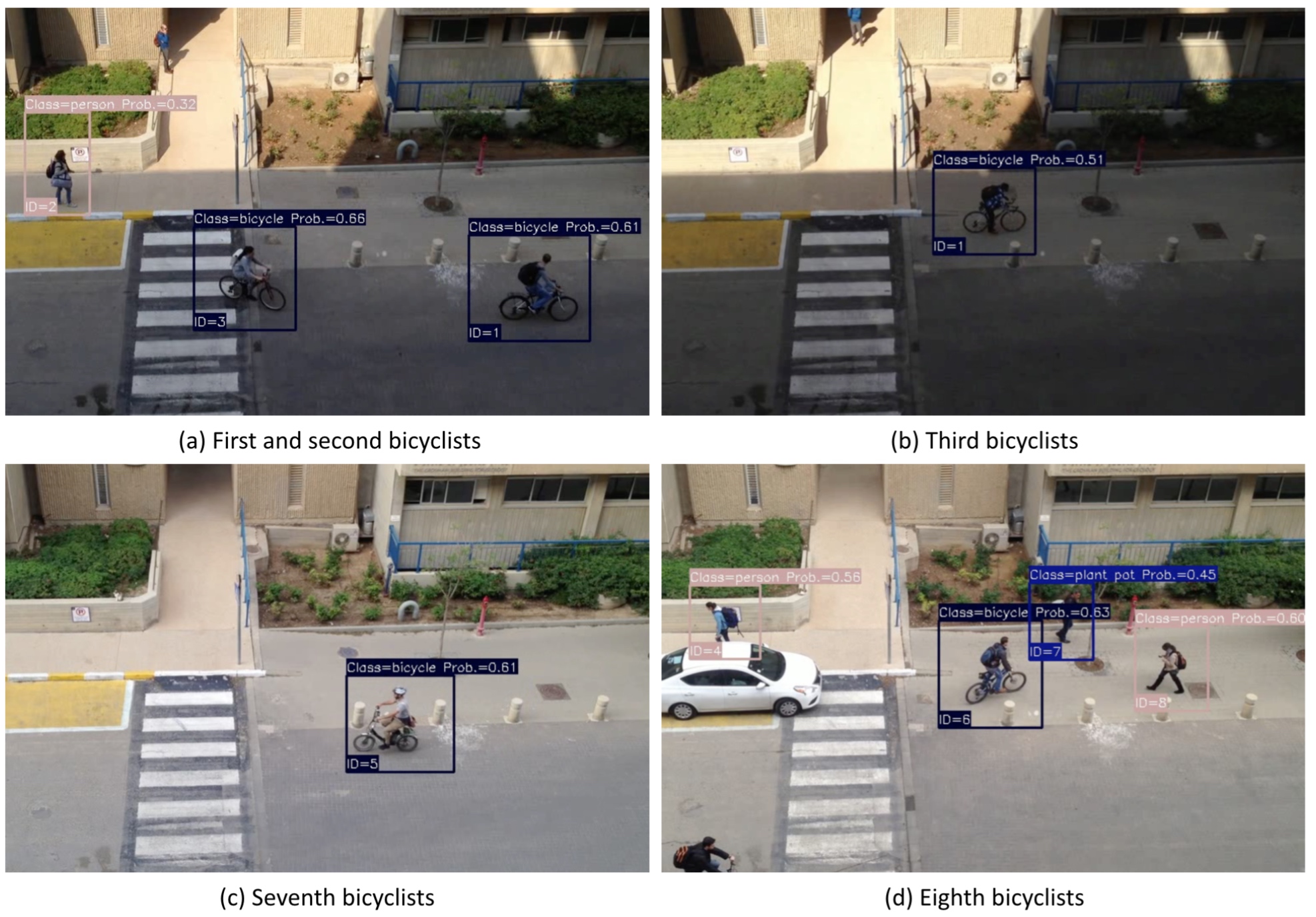}
\caption{Classification results of the bottom-up pathway (FRCNN with ZF model) for bicyclists passing through the test scene. Examining the confidence level of the system for the (a) first and second bicyclists, (b) the third bicyclist, (c) the seventh bicyclist, and (d) the eight bicyclist, shows constant performance. In addition, one of the people in (d) is misclassified as a plant due to the vegetation behind him.}
\label{fig:12}       
\end{figure}

The same frames when running the bottom-up pathway only can be seen in Fig. \ref{fig:12}. It is clear that the confidence levels for the bicyclists are approximately constant and are lower than those achieved by the VAP model. Furthermore, the classification of the people in the scene get a lower confidence level, and in subfigure (d) one person is also misclassified as a plant due to the background of vegetation behind him.

\begin{figure}
\centering
  \includegraphics[width=0.81\textwidth]{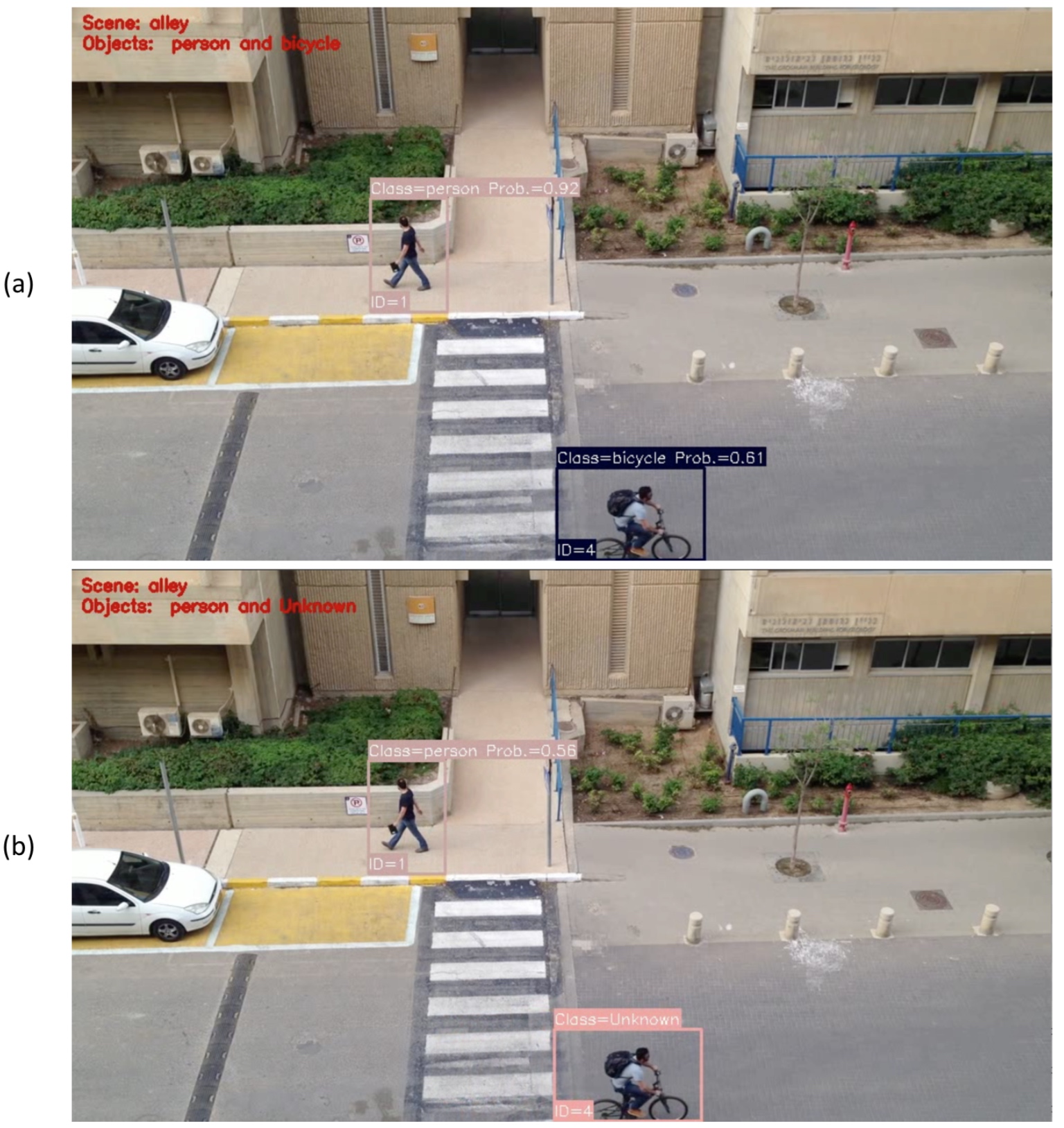}
\caption{Example of the VAP system operation. While the results of the VAP model, after watching the alley for about 30 minutes, resulted in high confidence recognition of the objects (a), using the same model without reinforcement learning failed to recognize one object and recognizes the other one with low confidence (b).}
\label{fig:13}       
\end{figure}

The strength of the reinforcement learning mechanism is evident when examining Fig. \ref{fig:13}. The bicycle rider naturally competes with the person class as it shares many similar features originating from the person riding the bicycle. As a result, the person in the scene is being correctly classified in both cases, but yields much higher confidence level with the full VAP model. The ability of the full VAP model to resolve this ambiguity is even more prominent for the bicycle rider, as the full VAP model correctly classifies him, while the non-learning model yields a very low confidence level that results with no recognition decision.

\newpage

Finally, Table \ref{tab:1} quantifies the performance comparison between the VAP model and the \acrshort{frcnn} algorithm. The values in the table are the accumulated mean average precision (mAP) and error rate of the system, after running the VAP model on all video clips in the dataset. These results validate the hypothesis underlying this work and show that including biologically-inspired mechanisms in computer vision systems would yield more robust and better performing object recognition architectures.

\begin{table}[hb]
\caption{Performance comparison between the faster RCNN and the VAP model.}
\label{tab:1}       
\centering
\begin{tabular}{lllll}
\hline\noalign{\smallskip}
CNN & FRCNN & FRCNN & VAP model & VAP model \\
 & [mAP] & [Error rate] & [mAP] & [Error rate] \\
\noalign{\smallskip}\hline\noalign{\smallskip}
ZF & 64.3\% & 0.32 & 93.0\% & 0.03 \\
VGG1024 & 68.4\% & 0.31 & 95.7\% & 0.03 \\
VGG16 & 82.5\% & 0.17 & 95.9\% & 0.03 \\
\noalign{\smallskip}\hline
\end{tabular}
\end{table}

\pagebreak

\section{Discussion}
\label{discussion}
The VAP model described in this work is proposed as a starting point for interaction between the computer vision and brain science communities and the development of synergistic models of artificial and biological vision. As the dynamics of neural processing is much more complex than the proposed computational model, lateral and recurrent interactions between the different levels of the model should be added to the VAP model. Such additions can improve the resemblance to the human visual system and can lead to a computational model that takes into account important aspects of human perception, such as detail preservation and active vision. A first step of adding lateral and recurrent interactions between the different levels of the model can be made by adding to the bottom-up CNN lateral interactions, such as in the U-Net \acrshort{cnn} \cite{Ref40} and recurrent interactions, such as in the Recurrent \acrshort{cnn} \cite{Ref41}.

The results using the VAP model show that the proposed method solves ambiguous classification cases better than bottom-up based algorithms, as well as achieving better classification performance throughout its operation. Nevertheless, one can argue that using a biologically-inspired predictive mechanism may result in false recognition, the same way the human visual perception is limited, as can be seen by numerous optical illusions. In addition, continual learning poses particular challenges for artificial neural networks due to the tendency for previously learnt knowledge to be abruptly lost as new information is incorporated \cite{Ref44}. This phenomenon, termed catastrophic forgetting, is specifically relevant for the VAP model as the model is trained on a specific dataset and then continuously trained on a different scene. Analysing and clarifying these effects and modifying the model in order to cope with them, is left for future work. 

Furthermore, keeping the original line of thinking of imitating the human visual system it can be interesting to adapt the VAP model in order to examine the hypothesis that predictions play not only a modulatory but also a driving role in awareness. A human observer can get a task to look for a specific type of a car in an entrance to a facility or counting the number of people entering the gate. For a human observer, these two different tasks will result in different priming and search strategies. The human eye-movement patterns, cognitive state and thought processes are systematically modulated by top down task demands \cite{Ref42,Ref43}. In order to add such mechanism to the VAP model, a new task/mission input should be introduced to the model, as well as new modules that will enable top-down high-level cognitive factors, such as task input, to modulate the the different parts of the model. Such a “mission control” management building block should lead to performance enhancement in the specific task, with a reduction in performance for competing visual inputs.

\section{Conclusions}
\label{conclusions}
Recent progress in understanding the way the human brain processes visual stimuli should lead to the development of artificial computer vision systems that emulate the human visual system. Such an approach is important not only for the computer vision community, for tasks where a human-like performance is required, but also for the brain research community, due to its potential to provide a tool for testing and evaluating the success in understanding the human visual system and its mechanisms.

The proposed model investigates the argument arising from recent brain research experiments which suggested that when visual input is ambiguous, predictions may help in the decision process and in maintaining a coherent interpretation of the environment. The VAP model imitates this expectation-driven perception mechanism, which is behind the recognition process in the human brain. The results using the model show that the proposed method solves ambiguous classification cases better than bottom-up based algorithms, while sustaining high classification performance throughout its operation.

In addition, state of the art results with deep neural networks are largely dependent on access to large labeled datasets relevant for the intended task. In narrow and unique domains for which large labeled datasets don't currently exist and the initial data set is small, an ongoing visual reinforcement learning can enable the possibility of exploiting the unsupervised stream of visual data that the unmanned agent receives during its operation, for improving its recognition abilities with time. In the proposed model, predictions play a driving role in a learning process that feeds a reinforcement learning mechanism. The results achieved with the model, show that this mechanism can enable a robotic agent to improve its recognition abilities with time by means of novel exploration and reinforcement learning. This could potentially also drive even better performing computer vision neural networks for tasks with large datasets, since they will be benefiting from models trained on gradually increasing datasets in the relevant context.

Overall, the results strongly support the integration of active top-down inferential process and reinforcement learning in visual perception algorithms. In addition, it implies that in the same way animals and humans actively acquire information about their environment by selecting sensory targets and probing their features, a robotic platform should employ the same tactic and use the continuous input of unlabeled images of its video stream for unsupervised learning, in order to discover and gain knowledge that wasn't available in the labeled data it was trained with.

Creating a computer vision algorithm that is massively inspired by the human visual system can be a useful tool in the brain research field as well as in the computer vision field. However, the human visual system has not been fully mapped yet and this gap in knowledge translates also to our inability to accurately model this system. Even for mechanisms that were well-researched, such as the mechanisms that are the basis of this work, there is a wide gap between our understandings of their general principles and the actual biological mechanisms that are dynamic in nature and based on multi-level excitatory and inhibitory interactions. Nevertheless, although the task of building a computational model of the human visual system is extremely complex and far from being accomplished, the potential benefits of developing such a model exceed the field of computer vision. It is important to note that there are inherent limitations for applying general principles of a biological mechanism, while not imitating all of the actual complex interactions in the human visual system. Nonetheless, a mutual multidisciplinary work on developing such a model can be used as a common ground for researchers from both fields to cooperate and develop a computational model of the human visual system, for the benefit of both fields of research.

\printglossary[type=\acronymtype]

\newpage

\bibliographystyle{unsrt}  


\end{document}